# A new method for protein structure reconstruction from NOESY distances

Z. Li, S. Li, X. Wei, X. Peng*, Q. Zhao*

ABSTRACT    Protein structure reconstruction from Nuclear Magnetic Resonance (NMR) experiments largely relies on computational algorithms. Recently, some effective low-rank matrix completion (MC) methods, such as ASD and ScaledASD, have been successfully applied to image processing, which inspires us to apply the methods to reconstruct protein structures. In this paper, we present an efficient method to determine protein structures based on experimental NMR NOESY distances. ScaledASD algorithm is used in the method with several post-procedures including chirality refinement, distance lower (upper) bound refinement, force field-based energy minimization (EM) and water refinement. By comparing several metrics in the conformation evaluation on our results with Protein Data Bank (PDB) structures, we conclude that our method is consistent with the popularly used methods. In particular, our results show higher validities in Procheck dihedral angles G-factor. Furthermore, we compare our calculation results with PDB structures by examining the structural similarity to X-ray crystallographic structures in a special dataset. The software and its MATLAB source codes are available in https://github.com/xubiaopeng/PRASD.

## INTRODUCTION

The characterization of three-dimensional protein structures has been a topic of great interest for decades since the knowledge of the protein structures is essential to understand the protein functions, which could lead to further pharmaceutical and medical applications. Currently, X-ray crystallography and nuclear magnetic resonance (NMR) spectroscopy (1) are two major experimental methods for protein structure determination. Unlike X-ray crystallography, NMR spectroscopy is not a "microscope with atomic resolution", but rather provides a network of distance measurements between spatially proximate hydrogen atoms (2); however, the introduction of NMR spectroscopy to characterize protein structure is a breakthrough because the method enables the identification of the protein structure in the aqueous solution, which is closer to the states of the proteins in cells. Since the first NMR-determined protein structure was reported in 1985 (1), thousands of protein structures have been measured by NMR spectroscopy, which are available in the Protein Data Bank (PDB) (3).

The typical NMR structure determination approach includes the following steps: peak picking from NMR spectra, chemical shift assignment (spectral assignment), geometric restraint assignment, and structural calculation (4). An important parameter in NMR experiments is Nuclear Overhauser Effect (NOE), which is used to generate distance restraints because the intensity of the NOE signal depends on the inverse sixth power of the internuclear distance (5). The NOE intensity between two atoms is reflected in the Nuclear Overhauser Effect Spectroscopy (NOESY), whose assignment relies on the knowledge of chemical shifts of nuclei. However, not all H-H distances can be measured accurately in the NMR experiments because the peak intensities corresponding to the long distances (larger than 5 Å) are too weak to be distinguished from the noise level in the experiment. Thus, usually just a network of short distances is available in the protein NMR structure measurement.

Since NMR measurements only provide implicit information about the protein structure, complicated computational algorithms have to be used to determine the protein structure in NMR experiment. Molecular dynamics (MD) incorporated with simulated annealing (SA) algorithm is one of the most widely used methods in protein NMR structure determination. MD (6) simulation is based on classical mechanics and often used to study the folding pathway and structure prediction (7-9), while SA (10) is a heuristic global optimization method to find the conformation at energy minimum. The SA procedure is usually not very fast. The performance of SA in different NP-hard problems was evaluated by Johnson *et al*. (11,12). MD/SA hybrid methods of XPLOR (13,14), DYANA (15), CYANA (16) and ARIA (17) were also implemented. The XPLOR is a method built on the MD simulation package CHARMM (18). The method can be used to search the conformations in Cartesian coordinate space. In contrast, DYANA and its improved version CYANA work in torsion angle space and hence have better calculation speed. ARIA can work in either torsion angle space or Cartesian coordinate space, but the algorithms are optimized for ambiguous distance restraints and violation analysis.

Euclidean distance matrix completion (EDMC) method is a different protein reconstruction method which mainly relies on the distance matrix obtained from NMR measurements. Compared to the MD/SA method, EDMC method uses fewer assumptions in energy function. The pioneer work of protein NMR by EDMC was done by Braun *et al*. in 1981 (19). Other efficient ways, such as EMBED (20) and DISGEO (21), were developed by Havel's group in 1983 and 1984, respectively. Later, semidefinite programming (SDP) was used to solve the EDMC problems by using Gram matrix instead of distance matrix, including DAFGAL proposed by Biswas *et al*. (22), DISCO proposed by Leung and Toh (23) and SPROS by Alipanahi *et al*. (24). Here we introduce a new matrix completion algorithm for protein structure reconstruction from NOESY distance restraints. By comparing our reconstructed results with PDB structures, we demonstrate that the method is valid and comparable with existing methods.

The paper is organized as follows: In Section 2, we describe our method for the NMR protein structure reconstruction, including the establishment of the initial matrix, matrix completion, distance bound refinement, chirality refinements, energy minimization (EM) optimization and water refinement. In Section 3, we test the validity of the method by comparing our calculation results with the PDB structures from several different aspects. Finally, the conclusion and future work are presented in Section 4.

# METHODS

Similar to the method for image recovery, the protein structure can be expressed as a distance matrix in the protein structure reconstruction, where each element is the distance of a pair of atoms. In this distance matrix, some elements can be known from the chemical properties, such as covalent bonds, and the NOESY distance restraints. On the other hand, additional refinements including the refinements on chirality and energetic stability in solution have to be applied to the recovered distance matrix due to the special properties of the protein structure. Based on the above considerations, we render the ScaledASD matrix completion algorithm, a successful algorithm in image processing, to determine the distance matrix from NOESY data, followed by further procedures to refine the structure.

## Initial distance matrix establishment

To determine the protein structure using the matrix completion, firstly we need to establish the initial distance matrix. The nonzero elements of this matrix stand for the known atomic distances. There are two types of known distances that can be directly filled into the matrix: 1) the distances measured in the NMR experiment, i.e., the distance restraints from NOESY data; 2) the distances between the coplanar atoms and all the covalent bonds. Although the coplanar and covalent bond distances can be varied with different residues, for a specific residue type, they can be considered as constants since the fluctuations are very small (25, 26).

However, a study has shown that it is insufficient to determine the structure uniquely with the above known distances (27). It was proposed in the study that the lower bound of the sample number $m$ in matrix completion theory should satisfy $m \geq Cnr\log_{10}n$, where $n$ and $r$ are the dimension and the rank of the matrix, respectively. The value of constant C is a certain positive number that is not exactly known. In our protein distance matrix completion, we can estimate this number by trials and tests. In detail, starting from a distance matrix of a protein with known structure, we sample $Cnr\log_{10}n$ elements uniformly and reconstruct the distance matrix using these samplings, where C is tested with different values from 0.5 to 5. For each C value we run recovery procedure ten times and calculate the average RMSD. The relation between the RMSD and C value is shown in Fig. S1 in Supplemental Information, indicating that the protein structure (distance matrix) is well recovered when C is larger than 2. We take the lower limit C = 2. For a typical NMR protein with 76 residues, such as 1G6J, atom number n~1200 and rank r = 5 for Euclidean Distance Matrix (EDM) (28), we need at least m~36000 to rebuild the matrix uniquely. In contrast, the NOESY data from NMR has only 1291 distances.

Together with all the covalent bonds and coplanar distances, there are 13488 distances in total, which is still far smaller to meet the matrix completion requirement.

To solve this problem, we add some distance elements estimated by the triangle inequality (29). In detail, assuming that there are N atoms named $A_1, A_2, A_3…A_N$, the boundary of an unknown distance $A_iA_j$ can be obtained from the known distances by the following conditions: $Max(|A_iA_k - A_kA_j|) \leq A_iA_j \leq Min(A_iA_k + A_kA_j)$ (for all k), where $A_k$ includes all the atoms whose distances $A_iA_k$ and $A_kA_j$ are known. In this paper, we just take the upper boundary as the corresponding element in the initial distance matrix for simplicity. Once the distance $A_iA_j$ is known in this way, we repeat this procedure including $A_iA_j$ as the known elements to determine other unknown distances until all the distance matrix elements are determined. Finally, the initial distance matrix is established by putting the following three types of distances together as described in Fig. 1: all the experimental NOESY data, all the covalent/coplanar distances and a sampled subset of the distances estimated from the triangle inequality.

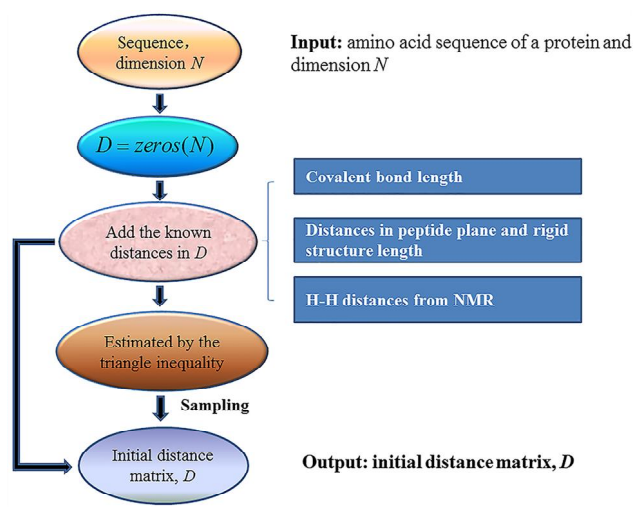

FIGURE 1    The process of establishing an initial distance matrix.

## Low-rank matrix completion

Matrix Completion (MC) (28, 30) problems are usually solved by minimizing the rank of matrices or minimizing the nuclear norm. However, both methods are time consuming (27,31). Recently, effective methods of alternating steepest descent (ASD) and scaled variant ScaledASD (32) are reported to solve the matrix completion problem directly with specified rank of matrix rather than using the nuclear norm minimization. These approaches were used to update the solutions by incorporating an exact line-search based on the factorization of the variable. ScaledASD is an accelerated version of ASD algorithm (32). In this paper, we only focus on ScaledASD. Supposing we have a matrix $Z_0$ with some elements are known in advance and the complete matrix Z can be factorized into two parts X and Y, i.e., Z=XY. The ScaledASD algorithm can find the proper factors X and Y so that the corresponding elements in matrix Z (Z=XY) are closest to the pre-known elements in $Z_0$. The flow chart of ScaledASD algorithm is shown in Fig. 2, where f(X,Y)=1/2‖$P_\Omega(Z_0)$-$P_\Omega(XY)$‖$_F$ stands for the Frobenius norm of the matrix $P_\Omega(Z_0)$-$P_\Omega(XY)$ and $P_\Omega$ is a sampling operator feching the known elements. The optimization loop stops when the solutions get converged, and we get the completed matrix at the end.

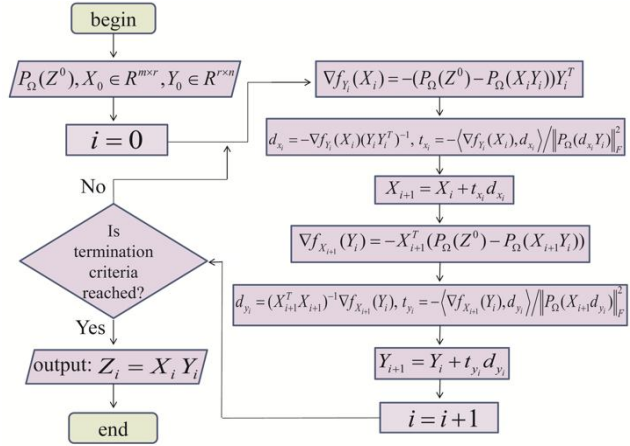

FIGURE 2    The flow chart of ScaledASD algorithm.

Once the distance matrix is recovered, we can easily reconstruct the coordinates from the Gram matrix following the reference (33).

## Distance bound refinement

In fact, the distance estimation by triangle inequality introduces certain errors into the initial distance matrix, which leads to inaccuracy in the structure calculation. Meanwhile, because the presence of the internal motions and chemical exchange may diminish the strength of the NOE signal (2), the interatomic distances obtained by NOE experiment tend to be larger than the actual distances. Therefore, we perform a distance bound refinement on the raw structure obtained from matrix completion to reduce the above errors. For this purpose, we take the refinement procedure used in reference (24), i.e., using nonlinear unconstrained optimization method BFGS (34) to minimize the following function

$$\phi(X) = \omega_E \sum_{(i,j)\in E} \left(\|x_i - x_j\| - e_{ij}\right)^2 + \omega_U \sum_{(i,j)\in U} f\left(\|x_i - x_j\| - u_{ij}\right)^2 \\ + \omega_L \sum_{(i,j)\in L} g\left(\|x_i - x_j\| - l_{ij}\right)^2 + \omega_R \sum_{i=1} \|x_i\|^2, \quad (1)$$

where $f(\alpha) = \max(0, \alpha)$ and $g(\alpha) = \min(0, -\alpha)$. The sets E, U, and L represent equality constraints, upper bounds and lower bounds, respectively; and $\omega_E, \omega_U, \omega_L$ are their relevant coefficients. We take the same setting for parameters $\omega_E = 2$, $\omega_U = 1$, $\omega_L = 1$ and

$$\omega_R = \alpha \frac{\phi(X^{(0)})|_{\omega_R=0}}{25\,\phi(X^{(0)})|_{\omega_R=1,\omega_E=\omega_U=\omega_L=0}} = \alpha \frac{1}{25 \sum_{i=1}^{n} \left\| x_i^{(0)} \right\|^2}. \quad (2)$$

where $-1 \leq \alpha \leq 1$ is a parameter.

Similar to the procedures in reference (24), we set the covalent bond lengths and distances among atoms on peptide plane as the equality constraints and a fraction of the sum of the van der Waals radii of atoms between non-bonded atom pairs as the lower bound to the corresponding distances:

$$d_{ij} \geq \sigma (r_i^{vdw} + r_j^{vdw}), \quad (3)$$

where $\sigma \in [0,1]$. $\sigma$ is typically around 0.85 (35). The values of the van der Waals radii are given in Table 1 (36, 37). However, we have different upper bounds in our algorithm: we include both the measured data from the NOE experiment and the estimated values from the triangle inequality as the upper bounds. As a result, by minimizing Eq.(1) starting from the rebuilt structure in Section 2.2, we actually make a complement on the upper limit approximation in both NOESY distance restraints and the triangle inequality estimation in the initial distance matrix establishment stage.

**TABLE 1    Van der Waals radii for different atoms**

| Atom | O | H | C | N |
|---|---|---|---|---|
| $r^{vdw}$ (Å) | 1.4 | 1.0 | 1.7 | 1.5 |

## Chirality refinements

Chirality is a geometric property of some molecules with mirror symmetry. As shown in Fig. 3, although molecules *A* and *B* have the same distance matrix, they are structurally different. In fact, the distance matrix does not provide any information about the global chirality, and hence the chirality refinement has to be performed additionally after the structure has been reconstructed from the distance matrix. Here, we use the same chirality refinement method as in reference (24). In detail, two types of chirality are considered: 1) The chirality of each residue: the residues in biological protein are mostly left handed (L-type) as shown in Fig.3 *B*. If the chirality of a residue is right handed, we just simply exchange positions of groups COOH and NH$_2$. 2) The chirality of the Ramachandran angle $\Phi$: Most of the $\Phi$ values should be negative. If the number of $\Phi \geq 0$ is larger than the number of $\Phi \leq 0$, we flip the structure by taking the opposite sign of all the x-components of the coordinates.

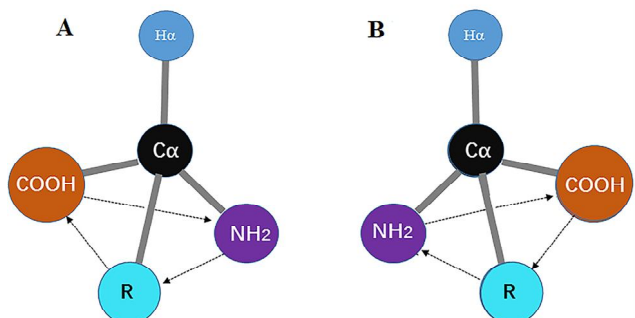

FIGURE 3  Schematic diagrams of chiral molecular structures. (A) stands for the incorrect enantiomer and (B) stands for the correct one in amino acids.

## EM optimization and water refinement

The protein structure has to be energetically stable in solution. For this requirement, we perform a force field-based EM optimization followed by water refinement on our reconstructed structure after the above geometrical refinements. The purpose of EM optimization is to get rid of the non-physical bonds and angles，and to find the configuration corresponding to the local minimum in energy landscape. The TIP3P water model and the AMBER99SB-ILDN (38) force field-based energy functions are used in the calculation. With the EM optimized structure, we carry out the process of thin-layer water refinement (39) where a simple annealing simulation on energy is used to obtain the energetically favored structure in solution.

## RESULTS AND DISCUSSION

We randomly picked ten proteins with different sizes and topologies measured by NMR in the PDB references as listed in Table 2. Table 2 also lists the most reliably well-defined regions by checking the NMR experimental report for each protein in PDB. For the input, we extract the NOESY distances from DOCR database in the NMR Restraints Grid (40,41), which is just a well-parsed database based on the NMR experimental data file in PDB. Then, we do a sampling from the distances estimated from the triangle inequality as described in Sec 2.1 at the ratio of $50/n$, where $n$ is the number of atoms. The sampled elements are input into the initial distance matrix. Recalling that the rank of a Euclidean distance matrix consistent with a protein structure is only 5, we notice that under current sampling ratio the lower bound requirement for matrix completion is satisfied for any protein with the number of atoms less than 10,000. The distance matrix of protein structure is recovered with ScaledASD method, followed by the geometrical refinements, EM optimization and finally the water refinement. We notice that the EM optimization could be force field dependent. However, as we tested with three different force fields -- AMBER99SB-ILDN (38), CHARMM36m (42) and OPLSAA (43), the reconstructed structure is robust to the force field selection. In the Supplemental Information, Table S1 shows the Cα RMSD in the well-defined region between structures obtained from different force fields. Therefore, we always use AMBER99SB-ILDN force field in our EM optimization stage in the reconstruction.

Clearly, our reconstructed structure is not unique due to the random sampling in establishing the initial distance matrix. For each protein, we repeat our reconstruction 20 times. It is found that the average Cα RMSD in well-defined region between two arbitrary structures from different runs is in the range 0.1~0.5Å, depending on the number of atoms, showing clear convergence. All our tests were completed on a desktop computer with a 3.4GHz processor and 8GB RAM. The matrix completion and structural refinements were performed with the software MATLAB 2016a; EM optimization and the water refinement were performed with GROMACS (44) and XPLOR (13), respectively. Depending on the sizes of the proteins in the test, the computation time for all our calculations varies from 10 minutes to 1.5 hours.

TABLE 2  Information about ten test proteins

| PDB ID | Description | Topology | Atoms | Residues | Well-defined region in PDB | Software |
|---|---|---|---|---|---|---|
| 1G6J | Ubiquitin | α+β | 1228 | 76 | 1-70 | DYANA |
| 1B4R | PKD domain 1 | β | 1114 | 80 | 8-87 | XPLOR |
| 2K62 | Liver fatty acid-binding protein | β | 1267 | 125 | 2-125 | CYANA |
| 1CN7 | Yeast ribosomal protein | α/β | 1648 | 104 | 9-70; 77-81; 89-101 | NMRDRAW; NMRPIPE |
| 2K49 | UPF0339 protein SO3888 | α+β | 1823 | 118 | 3-111 | AutoStructure; CYANA |
| 2L3O | Murine interleukin-3 | α | 1980 | 127 | 43-138 | CYANA; XPLOR |
| 2GJY | Tensin 1 PTB Domain | α+β | 2196 | 144 | 5-61; 67-87; 95-104; 111-137 | DYANA |
| 2K7H | Stress-induced proteinSAM22 | α/β | 2337 | 157 | 2-157 | XPLOR |
| 2YT0 | Amyloid beta A4 protein | α+β | 2602 | 176 | 18-30; 61-176 | CYANA |
| 2L7B | Apolipoprotein E | α | 4792 | 307 | 3-181; 186-200; 209-281;287-296 | CYANA |

# Evaluation and comparison on reconstructed proteins

## RMSD on well-defined region and TM-score comparison

To evaluate the performance of our method, we calculate the Cα RMSD values between our reconstruction results and the first model in their corresponding PDB structures in the well-defined regions as listed in Table 2. There are eight out of ten proteins with RMSD < 2 Å, indicating that our reconstruction is generally successful.

For the two proteins with RMSD > 2 Å, one is just 2.08 Å, while the other is 2.21 Å but with more than 300 residues. Reference (45) has shown that RMSD is length dependent, i.e., larger protein tends to have larger RMSD with the same structural similarity. To properly compare the structural similarity for the protein pairs with different lengths, we employ a length independent quantity TM-score (45). TM-score is a number in the range (0,1], where 1 indicates a perfect match between two structures and 0 indicates a complete mismatch. Usually, a score higher than 0.5 indicates that the two structures have the same fold in SCOP/CATH (46). TM-scores between our calculations and PDB structures are listed in the third column in Table 3, showing very high similarity between them.

Finally, we show the corresponding protein structures in Fig. 4, where the red represents the first model structure of the PDB references, and the blue represents one model of our calculated structure.

**TABLE 3** **The Cα RMSD in well-defined region and TM-scores for ten proteins. The RMSD and TM scores are calculated between the reconstructed ensemble and the first model of the corresponding PDB structure. The standard deviation denotes the uncertainty of the reconstructed structures**

| PDB ID | RMSD /well-defined (Å) | TM-score |
| --- | --- | --- |
| 1G6J | 1.00±0.10 | 0.87±0.01 |
| 1B4R | 1.57±0.16 | 0.85±0.02 |
| 2K62 | 1.81±0.14 | 0.86±0.01 |
| 1CN7 | 1.53±0.13 | 0.79±0.02 |
| 2K49 | 1.52±0.10 | 0.86±0.02 |
| 2L3O | 2.08±0.19 | 0.68±0.01 |
| 2GJY | 1.63±0.09 | 0.80±0.01 |
| 2K7H | 1.85±0.22 | 0.91±0.02 |
| 2YT0 | 0.95±0.07 | 0.78±0.01 |
| 2L7B | 2.21±0.14 | 0.88±0.01 |

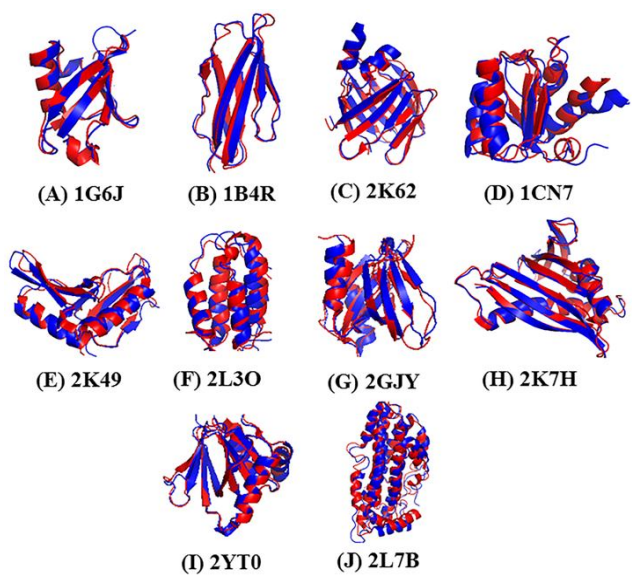

FIGURE 4  Superimposition of original PDB structures (red) and calculated structures (blue).

## Structure Validation using MPscore and PSVS

To further validate our reconstruction method, we utilize the Molprobity (47) and protein structure validation software suite (PSVS) (48) to evaluate the whole proteins reconstructed from our algorithm.

In Molprobity, there is a quantity MPscore for assessing the overall quality of the prediction structure from the statistical viewpoint. Lower values indicate better models. PSVS is another NMR standard validation tool to check both the geometric knowledge-based validation and the fit between the structure and the experimental data. For NMR structures, there are five geometric validation scores that are noteworthy: Molprobity clash-score (47), Procheck Phi-Psi and all dihedral angle G-factor (49), Verify3D score (50) and ProsaII score (51). In PSVS, they are all rescaled to their corresponding Z-scores. Usually a higher Z-score indicates a better model and a positive Z-score indicates that the analyzed structure is better than the typical high-resolution X-ray structure. PSVS report also gives the number of violations and maximum violation distance compared to the NOE experimental result. Here we normalize the violation number into violation percentage using the number of total restraints for comparison among proteins. Table 4 lists all these quantities in both our calculation results and the corresponding PDB entries for comparison. As shown in Table 4, our model generally performs better on Procheck Phi-Psi and all dihedral angle G-factor. On some other evaluation metrics such as ProsaII, MPscore and Molprobity clash-score, our calculation results are comparable with the corresponding PDB deposits. On violation percentage and the maximum violation distance, our calculation results are slightly worse than the PDB deposits.

**TABLE 4** MPscores, the Z-score and RMS of distance violation/constraint of test protein structures

| Protein ID | | MP-score | Verify 3D | ProsaII | Procheck (phi-psi/all) | MolProbity Clash-score | Viol./Cons. (%) | Max dist. viol. (Å) |
|---|---|---|---|---|---|---|---|---|
| 1G6J | Ref | 3.53 | -0.96 | 0.79 | -1.49/-4.55 | -5.77 | 0.49 | 0.41 |
| | Cal | 2.70 | -0.16 | 0.87 | -0.59/-1.30 | -2.27 | 5.38 | 2.58 |
| 1B4R | Ref | 4.39 | -1.93 | -1.94 | -4.45/-7.75 | -23.36 | 5.91 | 2.29 |
| | Cal | 3.53 | -1.77 | -1.82 | -3.03/-5.26 | -6.16 | 20.29 | 2.74 |
| 2K62 | Ref | 3.46 | -0.80 | -0.54 | -2.52/-4.26 | -1.79 | 36.82 | 7.06 |
| | Cal | 3.43 | -0.96 | -0.70 | -1.38/-3.43 | -8.05 | 26.18 | 3.22 |
| 1CN7 | Ref | 2.80 | 0.16 | 0.17 | -0.87/-2.60 | 0.45 | 2.71 | 1.23 |
| | Cal | 2.80 | -0.64 | -0.62 | -0.79/-1.95 | -2.09 | 10.45 | 2.71 |
| 2K49 | Ref | 2.11 | -2.41 | -1.41 | -0.87/-0.71 | -0.73 | 9.78 | 2.51 |
| | Cal | 2.56 | -2.25 | -1.36 | -0.55/-1.06 | -1.90 | 11.39 | 2.78 |
| 2L3O | Ref | 4.29 | -3.85 | -1.32 | -0.75/-5.32 | -11.70 | 17.72 | 3.65 |
| | Cal | 4.04 | -3.85 | -1.49 | 0.24/-2.90 | -11.07 | 24.73 | 4.16 |
| 2GJY | Ref | 3.71 | -1.44 | -0.79 | -2.28/-4.79 | -4.01 | 4.74 | 1.08 |
| | Cal | 3.29 | -2.89 | -1.28 | -1.93/-3.43 | -4.23 | 11.47 | 3.68 |
| 2K7H | Ref | 2.95 | -0.32 | -0.12 | -0.35/-0.59 | -11.53 | 14.48 | 4.73 |
| | Cal | 3.08 | 0.00 | -0.21 | -0.83/-2.07 | -4.14 | 16.19 | 2.77 |
| 2YT0 | Ref | 2.98 | -0.96 | 0.41 | -1.97/-3.43 | -2.67 | 22.33 | 2.07 |
| | Cal | 3.23 | -1.28 | 0.37 | -0.71/-1.42 | -3.24 | 23.65 | 2.27 |
| 2L7B | Ref | 3.82 | -3.05 | -0.95 | -0.47/-3.90 | -3.98 | 21.01 | 3.22 |
| | Cal | 4.13 | -3.21 | -1.41 | 0.63/-1.89 | -12.45 | 26.56 | 4.03 |

**Comparison on the protein structure properties**

In Sec 3.1, we have evaluated our results and compared them with the corresponding PDB structures using the scores provided in protein validation tools including Molprobity and PSVS. All of the results are in the overall evaluation scores. For most proteins, the local geometrical properties are sometimes more important in their biological functions. In this section, we evaluate our results on two local geometrical properties on backbone: the secondary structure and the Ramachandran Plot. In the following, we just take one arbitrary PDB structure 2K7H as an example. For the other proteins, similar results are obtained and listed in Supplemental Information.

**Secondary structure comparison**

Protein secondary structure is the 3D form of local segments, which is maintained by hydrogen bonds formed between carbonyl and amide groups on the skeleton. The secondary structure reflects the stability of protein structure. Since α-helices and β-strands are the most common secondary structures, we compared the regions of α-helices and β-strands in our calculated structures with the secondary structures in PDB references, respectively. The secondary structures are classified by DSSP algorithm (52). We select 2K7H as an example and show the secondary structures from our calculation results and the comparisons with the PDB structures in Fig. 5. Clearly, the secondary structures in the calculation results are almost the same as in PDB models except the very short segment 75-77, which is identified as loop in our calculation result but helix in the PDB structure. The similar comparisons for other proteins are shown in Fig. S2 in Supplemental Information.

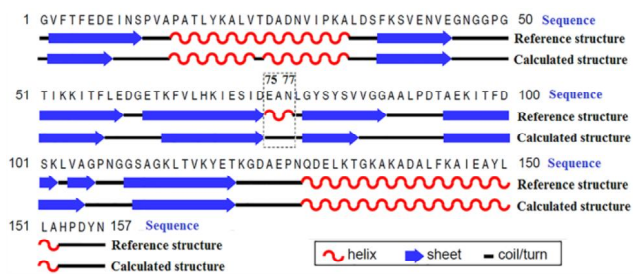

FIGURE 5  The secondary structure regions of 2K7H structure calculated based on ScaledASD and the PDB reference structure.

## Ramachandran Plot comparison

Ramachandran Plot is a way to visualize energetically allowed regions for backbone dihedral angles $\psi$ and $\phi$ of amino acid residues in protein structures (53). Usually there are three regions in Ramachandran Plot: favorable regions, allowed regions and disallowed regions. The combinations of ($\phi,\psi$) in favorable regions indicate that the results are without steric clashes; allowed regions indicate that the results are allowed even if the steric constraints are slightly relaxed; and the disallowed regions indicates that the results involve steric hindrance. We calculated the percentages of favourable and disallowed regions in the Ramachandran Plot of the calculated structures by MolProbity (47). The results of our calculation and the PDB structures are both listed in Table 5, where we can see that the percentages in favourable region of our calculated structures are comparable to the reference structures. To further confirm that the Ramachandran angles of our calculated structure are in the correct region, we compared the Ramachandran plots as shown in Fig. 6, where we can see the clear consistence between our calculation result and the PDB structure, except that there are two outliers in our calculation result.

TABLE 5  The percentages of the favourable and disallowed regions in Ramachandran Plot for the ten calculated protein structures as well as PDB references

| Protein ID | Favourable (reference) | Favourable (calculation) | Disallowed (reference) | Disallowed (calculation) |
|---|---|---|---|---|
| 1G6J | 92.23±2.40 | 94.66±2.30 | 0.63±0.91 | 1.01±1.51 |
| 1B4R | 83.14±1.87 | 81.80±3.85 | 3.91±1.14 | 6.73±1.71 |
| 2K62 | 87.69±2.46 | 87.80±1.94 | 0.93±0.56 | 2.40±1.66 |
| 1CN7 | 79.51±2.56 | 84.12±2.78 | 4.02±1.71 | 2.89±1.89 |
| 2K49 | 94.57±1.66 | 92.99±2.39 | 0.30±0.42 | 1.47±0.97 |
| 2L3O | 74.18±2.02 | 70.36±4.07 | 8.86±1.85 | 12.64±3.07 |
| 2GJY | 75.99±3.12 | 78.72±3.04 | 7.29±1.96 | 6.20±1.51 |
| 2K7H | 97.29±1.08 | 93.06±2.78 | 1.10±0.24 | 1.33±1.10 |
| 2YT0 | 85.55±2.76 | 79.67±2.39 | 1.15±0.56 | 6.07±1.36 |
| 2L7B | 75.69±1.58 | 73.38±1.92 | 6.41±1.19 | 9.11±1.61 |

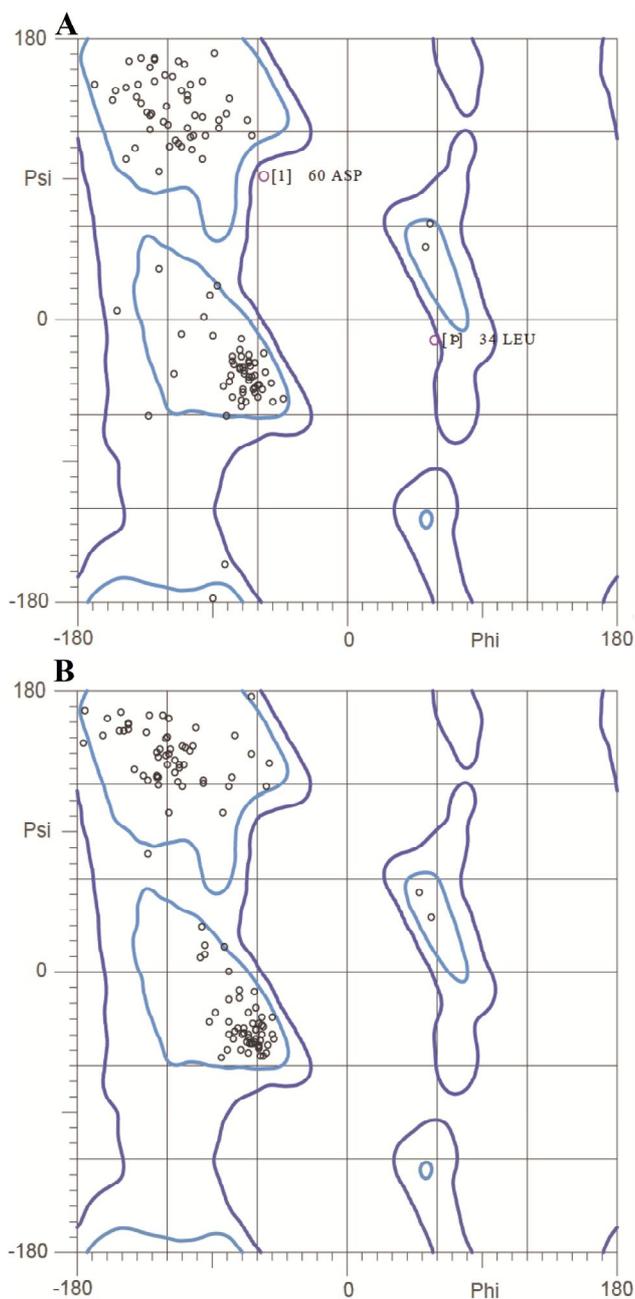

FIGURE 6  The Ramachandran Plot of 2K7H structures: (A) our calculation result; (B) the PDB reference structure.

## Validation using X-ray crystallographic structure

Finally, we validate our method using a special dataset, which has both NMR structures and the corresponding X-ray crystallographic structures in PDB. The dataset is set up as follows: Among the ten proteins that we used in Sec. 3.1, there are four proteins with their corresponding X-ray crystallographic structures. In addition, we take the PDB entries listed in references (54, 55), and remove the proteins whose RMSD values are larger than 3Å between the PDB NMR structures and the corresponding X-ray ones. In the end, we get a dataset with 31 proteins. We calculate the Cα RMSD values in the well-defined region between our calculated structure and X-ray crystallographic PDB structure. In the meantime, we also calculate the same Cα RMSD but for the PDB NMR structure compared with the X-ray crystallographic structure in PDB. The results are shown in Fig.7 *A*. For 14/31 proteins our structures are equivalent or even closer to the X-ray crystallographic structures than PDB NMR structures; for the rest, 9 of our structures are just slightly further (ΔRMSD<0.5Å) to the X-ray crystallographic structures than PDB NMR structures; for proteins with PDB code 1BVM and 1MPH, our reconstruction results are visibly worse than the PDB reference structures (ΔRMSD>1Å). We also compared the TM-score relative to the X-ray structure between our calculated structure and corresponding PDB reference structure, as shown in Fig.7 *B*. Similar to RMSD, the TM-scores of our structure are generally slightly smaller than the PDB structure, but all are larger than 0.65 indicating that the foldings are still the same as the X-ray structures. One possible reason that our calculated structure is slightly further to the X-ray crystallographic structure than the PDB reference structure could be due to the fact that we only consider the NOESY distance restraints in the NMR measurement while the PDB structures take all the possible NMR measurements into consideration during the reconstruction, which is the drawback of our current reconstruction algorithm.

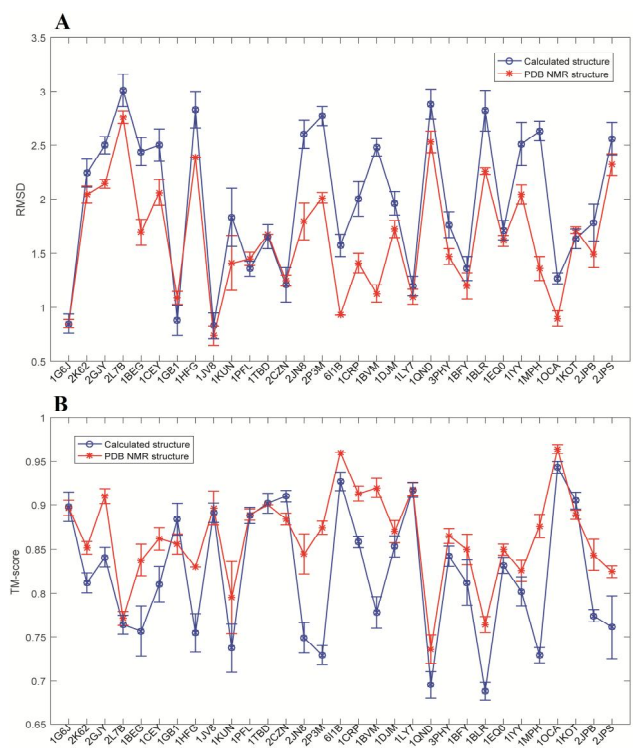

**FIGURE 7** (A) Cα RMSD in well-defined region relative to the X-ray crystallographic structure for both our calculation result (blue line) and PDB structure (red line). The bar indicates the standard deviation of the NMR ensemble, i.e., the reconstruction uncertainty. (B) Same as panel (A) but measured for TM-score.

## CONCLUSION

We have presented a new NMR structure determination method using ScaleASD matrix completion algorithm. The difficulties to reconstruct the NMR structure using matrix completion are from the following two facts: 1) The number of the measured NOESY distance restraints is far smaller than the required number in matrix completion condition. 2) The NOESY distances are usually larger than the actual atomic distances. To solve the first problem, we utilize the triangle inequality to generate the proximate values for some unknown distances. Using the samplings from these distances together with the measured NOESY data, covalent bonds and the coplanar inter-atomic distances, we set up the initial distance matrix that satisfies the matrix completion condition. For the second problem as well as the errors introduced by the triangle inequality estimation, we perform the distance boundaries refinement as complement. In the end, several refinements and optimization are implemented to obtain more reasonable structure. The software and the MATLAB source codes are distributed at https://github.com/xubiaopeng/PRASD. The release is archived in Zenodo: DOI: 10.5281/zenodo.1400047.

To validate our method, ten arbitrarily selected proteins from NOESY data have been reconstructed and compared with their corresponding structures in PDB. The comparison includes several metrics such as Cα atom RMSD in well-defined region, TM-score, MPscore, Z-scores and violations reported in PSVS software, secondary structures and Ramachandran plots. It has been shown that our results are comparable to the structures in PDB, especially for Procheck dihedral angles G-factor, the results from our method show higher validity than the PDB structures. We have also validated our reconstructed structure using the RMSD and TM-score with the corresponding PDB X-ray crystallographic structure with an even larger dataset. The result shows that our method is valid even though not as good as the current PDB NMR structures. We notice that our current reconstruction structure is only based on NOESY distance restraints. In the next step, many other NMR measurements such as Residual dipolar coupling (RDC) would be considered and integrated into our algorithm to obtain better structures.

## AUTHOR CONTRIBUTIONS

## ACKNOWLEDGMENTS

The authors would like to thank Prof. Ge Molin for valuable discussions, Prof. Deng Rongping for reading the manuscript and improving English. This work was supported by the National Science Foundation (NSF) of China with the Grant No.11675014. Additional support was provided by the Ministry of Science and Technology of China (2013YQ030595-3).